\documentclass[conference]{IEEEtran}

\usepackage{PerlazaISIT2025-V2}
\usepackage[utf8]{inputenc} 
\usepackage[T1]{fontenc}
\usepackage{url}
\usepackage{ifthen}
\usepackage{cite}
\usepackage[left=0.75in,right=0.75in,top=0.75in,bottom=1.1in]{geometry}

\begin{document}
\title{Variations on the Expectation \\ due to Changes in the Probability Measure} 
 \author{%
   \IEEEauthorblockN{Samir M. Perlaza\IEEEauthorrefmark{1}\IEEEauthorrefmark{2}\IEEEauthorrefmark{3} and 
                     Gaetan Bisson\IEEEauthorrefmark{3}}
   \IEEEauthorblockA{\IEEEauthorrefmark{1}%
                     INRIA, Centre Inria d'Université Côte d'Azur,
                     Sophia Antipolis, France.}
   \IEEEauthorblockA{\IEEEauthorrefmark{2}%
                     ECE Dept. Princeton University, 
                     Princeton, 08544 NJ, USA.}
   \IEEEauthorblockA{\IEEEauthorrefmark{3}%
		     Laboratoire GAATI, University of French Polynesia.
\thanks{This work is supported in part by the European Commission through the H2020-MSCA-RISE-2019 project 872172; the French National Agency for Research (ANR)  through the Project ANR-21-CE25-0013 and the project ANR-22-PEFT-0010 of the France 2030 program PEPR Réseaux du Futur; and in part by the Agence de l'innovation de défense (AID) through the project UK-FR 2024352. }                     } 
  }

%

\maketitle


\begin{abstract}
In this paper, closed-form expressions are presented for the variation of the expectation of a given function due to changes in the probability measure used for the expectation.  They unveil interesting connections with Gibbs probability measures,  mutual information, and lautum information.
\end{abstract}

\section{Introduction}

Let $m$ be a positive integer and denote by $\triangle(\reals^m)$ the set of
all probability measures on the measurable space $\left(\reals^{m},\BorSigma{\reals^{m}}\right)$, with $\BorSigma{\reals^{m}}$ being the Borel~$\sigma$-algebra on $\reals^m$. Given a Borel measurable
function $h:\reals^n\times\reals^m \to \reals$, consider the
functional~$\mathsf{G}_h: \reals^n\times\triangle(\reals^m) \times
\triangle(\reals^m) \to \reals$ such that
\begin{IEEEeqnarray}{lCl}
\label{EqDecember11at20h03in2024AgadirCity}
\mathsf{G}_h\left( x, P_1, P_2 \right) & = & \int h(x,y) \mathrm{d}P_1(y) - \int h(x,y) \mathrm{d}P_2(y),\IEEEeqnarraynumspace
\end{IEEEeqnarray}
which quantifies the variation of the expectation of the measurable function $h$ due to changing
the probability measure from $P_2$ to $P_1$. Such a functional is defined when both integrals exist and are finite. 

In order to define the expectation of $\mathsf{G}_h\left( x, P_1, P_2 \right)$
with respect to $x$, the structure formalized below is required.

\begin{definition}
A family $P_{Y|X} \triangleq (P_{Y|X=x})_{x\in\reals^n}$ of elements of
$\triangle(\reals^m)$ indexed by $\reals^n$ is said to be a conditional
probability measure if,
for all sets~$\set{A} \in \BorSigma{\reals^{m}}$, the map 
\begin{IEEEeqnarray}{cCl}
\nonumber
\function{\reals^n}{[0,1]}{x}{P_{Y|X=x}(\set{A})}
\end{IEEEeqnarray}
is Borel measurable.
The set of all such conditional probability measures 
is denoted by $\triangle\left(\reals^m | \reals^{n} \right)$.
\end{definition}

In this setting, consider the functional~$\bar{\mathsf{G}}_h: \triangle\left(\reals^m | \reals^{n} \right) \times  \triangle\left(\reals^m | \reals^{n} \right) \times \triangle\left(\reals^{n} \right) \to \reals$ such that
\begin{IEEEeqnarray}{lCl}
\label{EqDecember11at20h13in2024AgadirCity}
\bar{\mathsf{G}}_h\left( P_{Y|X}^{(1)}, P_{Y|X}^{(2)}, P_{X} \right)  & =  &\int \mathsf{G}_h\left( x, P_{Y|X=x}^{(1)}, P_{Y|X=x}^{(2)} \right)   \mathrm{d}P_{X}(x). \quad
\end{IEEEeqnarray}
This quantity can be interpreted as the variation of the integral (expectation) of the function $h$ when the probability measure changes from the joint probability measure $P_{Y|X}^{(1)} P_{X}$ to another joint probability measure $P_{Y|X}^{(2)} P_{X}$, both in $\triangle\left( \reals^{m} \times \reals^{n} \right)$. This follows from~\eqref{EqDecember11at20h13in2024AgadirCity} by observing that 
\begin{IEEEeqnarray}{rcl}
\nonumber
& & \bar{\mathsf{G}}_{h}\left( P_{Y|X}^{(1)}, P_{Y|X}^{(2)}, P_{X} \right) \\
& = & \int h(x,y) \mathrm{d}P_{Y|X}^{(1)} P_{X} (y,x) -   \int h(x,y) \mathrm{d}P_{Y|X}^{(2)} P_{X} (y,x).\IEEEeqnarraynumspace 
\end{IEEEeqnarray}

Special attention is given to the quantity $\bar{\mathsf{G}}_h\left( P_{Y}, P_{Y|X}, P_{X} \right)$, for some $P_{Y|X} \in \triangle\left(\reals^m | \reals^{n} \right)$,  with~$P_{Y}$ being the marginal of the joint probability measure~$P_{Y|X} \cdot P_{X}$. 
That is, for all sets $\set{A} \in \BorSigma{\reals^{m}}$,
\begin{IEEEeqnarray}{rCl}
\label{EqDecember25at20h14in2024Nice}
P_{Y}\left( \set{A} \right) & = & \int P_{Y|X = x}\left( \set{A} \right) \mathrm{d}P_{X}(x).
\end{IEEEeqnarray}
Its relevance stems from the fact that it captures the variation of the expectation of the function $h$ when the probability measure changes from the joint probability measure $P_{Y|X}P_{X}$ to the product of its marginals $P_{Y} P_{X}$. That is, 
\begin{IEEEeqnarray}{lcl}
\nonumber
& &\bar{\mathsf{G}}_h\left( P_{Y}, P_{Y|X}, P_{X} \right)  \\
\nonumber
& = &\int \left(\int h(x,y) \mathrm{d}P_{Y} (y) - \int h(x,y) \mathrm{d}P_{Y | X = x} (y) \right) \mathrm{d}P_{X}(x)\\
\label{EqDecember24at12h06in2024Nice}
& = & \int h(x,y) \mathrm{d}P_{Y}P_{X}  (y,x) - \int h(x,y) \mathrm{d}P_{Y|X}P_{X} (y,x).
\end{IEEEeqnarray}

\subsection{Novelty and Contributions}
This work makes two key contributions: First, it provides a closed-form expression for the variation~$\mathsf{G}_{h}\left( x, P_1, P_2 \right)$ in~\eqref{EqDecember11at20h03in2024AgadirCity} for a fixed~$x \in \reals^{n}$ and two arbitrary probability measures~$P_1$ and~$P_2$, formulated explicitly in terms of information-theoretic quantities. Second, it derives a closed-form expression for the expected variation~$\bar{\mathsf{G}}_{h}\left(P_{Y|X}^{(1)}, P_{Y|X}^{(2)}, P_{X} \right)$ in~\eqref{EqDecember11at20h13in2024AgadirCity}, again in terms of information measures, for  arbitrary conditional probability measures~$P_{Y|X}^{(1)}$, $P_{Y|X}^{(2)}$, and marginal measure~$P_{X}$.

A further contribution of this work is the derivation of specific closed-form expressions for $\bar{\mathsf{G}}_h\left( P_{Y}, P_{Y|X}, P_{X} \right)$ in~\eqref{EqDecember24at12h06in2024Nice}, which reveal deep connections with both mutual information~\cite{Shannon-1948a, Shannon-1948b} and lautum information~\cite{palomar2008lautum}. Notably, when $P_{Y|X}$ is a Gibbs conditional probability measure, this variation simplifies (up to a constant factor) to the sum of the mutual and lautum information induced by the joint distribution $P_{Y|X}P_{X}$.

Although these results were originally discovered in the analysis of generalization error of machine learning algorithms, see for instance~\cite{Perlaza-ISIT2023b, PerlazaTIT2024, zou2024Generalization, zouJSAIT2024, perlaza2024generalization}, where the function $h$ in~\eqref{EqDecember11at20h03in2024AgadirCity} was assumed to represent an empirical risk, this paper presents such results in a comprehensive and general setting that is no longer tied to such assumptions. This new general presentation not only unifies previously scattered insights but also makes the results applicable across a broad range of domains in which changes in the expectation due to variations of the underlying probability measures are relevant.

\subsection{Applications}
The study of the variation of the integral (expectation) of~$h$ (for some fixed~$x \in \reals^{n}$) due to a measure change from~$P_2$ to~$P_1$, i.e., the value~$\mathsf{G}_h\left( x, P_1, P_2 \right)$ in~\eqref{EqDecember11at20h03in2024AgadirCity}, plays a central role in the definition of \emph{integral probability metrics} (IPMs)\cite{muller1997integral, zolotarev1983probability}. 
Using the notation in~\eqref{EqDecember11at20h03in2024AgadirCity}, an IPM results from the optimization problem
\begin{IEEEeqnarray}{rcl} 
\sup_{h \in \set{H}} \abs{\mathsf{G}_h\left( x, P_1, P_2 \right)},
\end{IEEEeqnarray}
for some fixed~$x \in \reals^{n}$ and a particular class of functions~$\set{H}$.
Note for instance that the maximum mean discrepancy is an IPM  \cite{JMLR:v13:gretton12a}, as well as the Wasserstein distance of order one \cite{villani2009optimal, liu2024information, wenliang2024Out, JMLR:v22:20-867}.  

Other areas of mathematics in which the variation~$\mathsf{G}_h\left( x, P_1, P_2 \right)$ in~\eqref{EqDecember11at20h03in2024AgadirCity} plays a key role is distributionally robust optimization (DRO) \cite{rahimian2022frameworks,xu2024flow} and optimization with relative entropy regularization \cite{PerlazaTIT2024, zou2024Generalization}. In these areas, the variation~$\mathsf{G}_h\left( x, P_1, P_2 \right)$ is a central tool. See for instance, \cite{hu2013kullback, Perlaza-ISIT2023b}.
Variations of the form~$\mathsf{G}_h\left( x, P_1, P_2 \right)$ in~\eqref{EqDecember11at20h03in2024AgadirCity} have also been studied in \cite{zouJSAIT2024} and~\cite{perlaza2024generalization} in the particular case of statistical machine learning for the analysis of generalization error. The central observation is that   the generalization error of machine learning algorithms can be written in the form~$\bar{\mathsf{G}}_h\left( P_{Y}, P_{Y|X}, P_{X} \right)$ in \eqref{EqDecember24at12h06in2024Nice}. This observation is the main building block of the \emph{method of gaps} introduced in~\cite{perlaza2024generalization}, which leads to a number of closed-form expressions for the generalization error involving mutual information, lautum information, among other information measures. 

\section{Preliminaries}
\label{sec:i}
The main results presented in this work involve Gibbs conditional probability measures.
Such measures are parametrized by a Borel measurable function~$h: \reals^{n} \times \reals^{m}  \to \reals$; a~$
\sigma$-finite measure~$Q$ on $\reals^m$; and a vector~$x \in \reals^{n}$.
Note that the variable $x$ will remain inactive until Section~\ref{sec:iii}.
Although it is introduced now for consistency, it could be removed altogether from all results presented in this section and  Section~\ref{sec:ii}.

Denote by~$\mathsf{K}_{h, Q, x}: \reals \to \reals$ the function that satisfies
\begin{IEEEeqnarray}{rCl}
\label{EqDecember15at11h13in2024Nice}
\mathsf{K}_{h, Q, x}\left( t \right) & = & \log\left( \int \exp\left( t \, h(x, y) \right)\mathrm{d}Q\left( y \right) \right).
\end{IEEEeqnarray}
Under the assumption that~$Q$ is a probability measure, the function~$\mathsf{K}_{h, Q, x}$ in~\eqref{EqDecember15at11h13in2024Nice} is the cumulant  generating function of the random variable~$h(x, Y)$, for some fixed~$x \in \reals^{n}$ and~$Y \sim Q$. 
Using this notation, the definition of the Gibbs conditional probability measure is presented hereunder.
\begin{definition}[Gibbs Conditional Probability Measure]
Given a Borel measurable function~$h: \reals^{n} \times \reals^{m}  \to \reals$; a~$
\sigma$-finite measure~$Q$ on $\reals^m$; and a~$\lambda \in \reals$, the probability measure~$P^{(h, Q, \lambda)}_{Y | X} \in \triangle\left(\reals^m | \reals^{n} \right)$ is said to be an~$(h, Q, \lambda)$-Gibbs conditional probability measure if
\begin{IEEEeqnarray}{rCl}
\label{EqDecember18at16h25in2024BusFromSophiaToNice}
\mbox{$\forall x \in \reals^{n}$, } \mathsf{K}_{h, Q, x} \left( - \lambda \right) & < & + \infty; 
\end{IEEEeqnarray}
and for all~$(x,y) \in \reals^{n} \times \reals^{m}$,
\begin{IEEEeqnarray}{rCl}
\label{December15at16h31in2024}
\frac{\mathrm{d}P^{(h, Q, \lambda)}_{Y | X = x }}{\mathrm{d}Q} \left( y \right)& = & \exp\left( -  \lambda h\left( x, y \right) - \mathsf{K}_{h,Q, x} \left( -  \lambda\right) \right),
\end{IEEEeqnarray}
where the function~$\mathsf{K}_{h, Q, x}$ is defined in~\eqref{EqDecember15at11h13in2024Nice}.
\end{definition}
Note that, while $P_{Y|X}^{(h,Q,\lambda)}$ is an $(h, Q, \lambda)$-Gibbs conditional probability measure,  the measure $P_{Y|X=x}^{(h,Q,\lambda)}$, obtained by conditioning it upon a given vector $x\in\reals^n$, is referred to as an~$(h,
Q, \lambda)$-Gibbs probability measure.

The condition in~\eqref{EqDecember18at16h25in2024BusFromSophiaToNice} is easily met under certain assumptions. For instance, if~$h$ is a nonnegative function and $Q$ is a finite measure, then it holds for all $\lambda\in \left( -\infty, 0\right)$.
Let $\triangle_{Q}\left( \reals^{m} \right) \triangleq \left\lbrace P \in \triangle\left( \reals^{m} \right): P  \ll Q \right\rbrace$, with $P \ll Q$ standing for ``$P$ absolutely continuous with respect to $Q$''.
The relevance of~$(h, Q, \lambda)$-Gibbs probability measures relies on the fact that under some conditions, they are the unique solutions to problems of the form,
\begin{IEEEeqnarray}{rcl}
\label{EqDecember18at11h09in2024Sophia}
& &\min_{P \in  \triangle_Q(\reals^m) }  \int h(x,y) \mathrm{d}P(y) + \frac{1}{\lambda} \KL{P}{Q}, \mbox{ and }\\
\label{EqDecember18at11h09in2024SophiaA}
& &\max_{P \in  \triangle_Q(\reals^m) }  \int h(x,y) \mathrm{d}P(y) + \frac{1}{\lambda} \KL{P}{Q},
\end{IEEEeqnarray}
where~$\lambda \in \reals\setminus\lbrace 0 \rbrace$, $x \in \reals$, and $\KL{P}{Q}$ denotes the relative entropy (or KL divergence) of $P$ with respect to $Q$.
The connection between the optimization problem \eqref{EqDecember18at11h09in2024Sophia} and the Gibbs probability measure $P^{(h, Q, \lambda)}_{Y | X = x }$ in \eqref{December15at16h31in2024} appeared for the first time in \cite[Theorem~$3$]{PerlazaTIT2024}. Alternatively, the connection between the optimization problem \eqref{EqDecember18at11h09in2024SophiaA} and such a Gibbs probability measure  appeared first in~ \cite[Theorem~$1$]{zouJSAIT2024}. In both cases, the presentation was in the context of a statistical learning problem. 
A general and unified statement of these observations is presented hereunder.
\begin{lemma}\label{LemmaDecember19at14h56in2024Nice}
Assume that the optimization problem in~\eqref{EqDecember18at11h09in2024Sophia} (respectively, in~\eqref{EqDecember18at11h09in2024SophiaA}) admits a solution. Then, if~$\lambda > 0$ (respectively, if~$\lambda < 0$), 
the probability measure $P^{(h, Q, \lambda)}_{Y | X = x }$ in \eqref{December15at16h31in2024} is the unique solution.  
\end{lemma}
\begin{IEEEproof}
For the case in which $\lambda >0$, the proof follows the same approach as the proof of \cite[Theorem~$3$]{PerlazaTIT2024}. Alternatively,  for the case in which $\lambda < 0$, the proof follows along the lines of the proof of \cite[Theorem~$1$]{zouJSAIT2024}.
\end{IEEEproof}

The following lemma highlights a key property of~$(h, Q, \lambda)$-Gibbs conditional probability measures. A special case of this result was first introduced in \cite[Lemma 3]{zouJSAIT2024} for the case in which $\lambda < 0$. Nonetheless, it holds in general for all $\lambda \in \reals\setminus\lbrace 0 \rbrace$, as shown by the following lemma.
\begin{lemma}
Given an~$(h, Q, \lambda)$-Gibbs probability measure, denoted by~$P^{(h, Q, \lambda)}_{Y | X = x }$, with~$x \in \reals^n$, 
\begin{IEEEeqnarray}{rcl}
\label{EqDecember19at14h36in2024Nice}
& & - \frac{1}{\lambda} \mathsf{K}_{h,Q, x} \left( -  \lambda\right) 
=  \int h(x,y)\mathrm{d}Q\left( y \right) - \frac{1}{\lambda}  \KL{Q}{P^{(h, Q, \lambda)}_{Y | X = x }}\squeezeequ \IEEEeqnarraynumspace \\
\label{EqDecember19at14h35in2024Nice}
& = &  \int h(x,y)\mathrm{d}P^{(h, Q, \lambda)}_{Y | X = x } \left( y \right) + \frac{1}{\lambda}  \KL{P^{(h, Q, \lambda)}_{Y | X = x }}{Q}; \IEEEeqnarraynumspace  
\end{IEEEeqnarray}
moreover, if~$\lambda > 0$, 
\begin{IEEEeqnarray}{rcl}
\label{EqDecember19at14h31in2024Nice}
 - \frac{1}{\lambda} \mathsf{K}_{h,Q, x} \left( -  \lambda\right)  & = & \min_{P \in  \triangle_Q(\reals^m) }  \int h(x,y) \mathrm{d}P(y) + \frac{1}{\lambda} \KL{P}{Q};\IEEEeqnarraynumspace\\
 \nonumber
\end{IEEEeqnarray}
alternatively, if~$\lambda < 0$,
\begin{IEEEeqnarray}{rcl}
\label{EqDecember19at14h31in2024NiceA}
 - \frac{1}{\lambda} \mathsf{K}_{h,Q, x} \left( -  \lambda\right)  & = & \max_{P \in  \triangle_Q(\reals^m) }  \int h(x,y) \mathrm{d}P(y) + \frac{1}{\lambda} \KL{P}{Q}, \IEEEeqnarraynumspace
\end{IEEEeqnarray}
where the function~$\mathsf{K}_{h, Q, x}$ is defined in~\eqref{EqDecember15at11h13in2024Nice}.
\end{lemma}

\begin{IEEEproof}
The proof of~\eqref{EqDecember19at14h35in2024Nice} follows from taking the logarithm of both sides of~\eqref{December15at16h31in2024} and integrating with respect to~$P^{(h, Q, \lambda)}_{Y | X = x }$. As for the proof of~\eqref{EqDecember19at14h36in2024Nice}, it follows by noticing that for all~$(x,y) \in \reals^n \times \supp Q$, the Radon-Nikodym derivative~$\frac{\mathrm{d}P^{(h, Q, \lambda)}_{Y | X = x }}{\mathrm{d}Q} \left( y \right)$ in~\eqref{December15at16h31in2024} is strictly positive. Thus, 
$\frac{\mathrm{d}Q}{\mathrm{d}P^{(h, Q, \lambda)}_{Y | X = x }}\left( y \right)   =  \left( \frac{\mathrm{d}P^{(h, Q, \lambda)}_{Y | X = x }}{\mathrm{d}Q} \left( y \right)\right)^{-1}$.
Hence, taking the negative logarithm of both sides of~\eqref{December15at16h31in2024} and integrating with respect to~$Q$ leads to~\eqref{EqDecember19at14h36in2024Nice}.
Finally, the equalities in~\eqref{EqDecember19at14h31in2024Nice} and~\eqref{EqDecember19at14h31in2024NiceA} follow from Lemma~\ref{LemmaDecember19at14h56in2024Nice} and~\eqref{EqDecember19at14h35in2024Nice}.  
\end{IEEEproof}
The following lemma introduces the main building block of this work, which is a characterization of the variation~$\mathsf{G}_h\left( x, P, P^{(h, Q, \lambda)}_{Y | X = x } \right)$.
Such a result appeared for the first time in \cite[Theorem~1]{Perlaza-ISIT2023b} for the case in which $\lambda > 0$; and in \cite[Theorem 6]{zouJSAIT2024} for the case in which $\lambda < 0$, in different contexts of statistical machine learning.  
A general and unified statement of such results is presented hereunder. 

\begin{lemma}\label{LemmaDecember20at17h39in2024BusFromSophiaToNice}
Consider an~$(h, Q, \lambda)$-Gibbs  probability measure, denoted by~$P^{(h, Q, \lambda)}_{Y | X = x } \in \triangle\left( \reals^{m} \right)$, with~$\lambda \neq 0$ and~$x \in \reals$. For all~$P \in \triangle_{Q}\left( \reals^{m} \right)$, 
\begin{IEEEeqnarray}{rcl}
\nonumber
& & \mathsf{G}_h\left( x, P, P^{(h, Q, \lambda)}_{Y | X = x } \right) \\
\label{EqDecember20at17h39in2024BusFromSophiaToNice}
& = & \frac{1}{\lambda} \left( \KL{P}{P^{(h, Q, \lambda)}_{Y | X = x }} + \KL{P^{(h, Q, \lambda)}_{Y | X = x }}{Q} - \KL{P}{Q} \right).
 \IEEEeqnarraynumspace
\end{IEEEeqnarray}
\end{lemma}
\begin{IEEEproof}
The proof follows along the lines of the proofs of \cite[Theorem~1]{Perlaza-ISIT2023b} for the case in which $\lambda > 0$; and in \cite[Theorem 6]{zouJSAIT2024} for the case in which $\lambda < 0$.
A unified proof is presented hereunder by noticing that for all~$P \in \triangle_{Q}\left( \reals^{m} \right)$, 
\begin{IEEEeqnarray}{rcl}
\nonumber
& & \KL{P}{P^{(h, Q, \lambda)}_{Y | X = x }} \\
\label{EqDecember22at19h48in2024}
&  = & \int  \log\left( \frac{\mathrm{d}P}{\mathrm{d}P^{(h, Q, \lambda)}_{Y | X = x }}  (y) \right) \mathrm{d}P(y) \\
\label{EqDecember22at19h48in2024A}
& = & \int  \log\left( \frac{\mathrm{d}Q}{\mathrm{d}P^{(h, Q, \lambda)}_{Y | X = x }}  (y)  \frac{\mathrm{d}P}{\mathrm{d}Q}  (y)  \right) \mathrm{d}P(y) \\
\label{EqDecember22at19h48in2024B}
& = & \int  \log\left( \frac{\mathrm{d}Q}{\mathrm{d}P^{(h, Q, \lambda)}_{Y | X = x }}  (y) \right) \mathrm{d}P(y) + \KL{P}{Q}\\
\label{EqDecember22at19h48in2024C}
& = & \lambda \int h(x,y) \mathrm{d}P(y) +  \mathsf{K}_{h,Q, x} \left( -  \lambda\right) + \KL{P}{Q}\\
\label{EqDecember22at19h48in2024D}
& = & \lambda \mathsf{G}_h\left( x, P, P^{(h, Q, \lambda)}_{Y | X = x } \right) - \KL{P^{(h, Q, \lambda)}_{Y | X = x }}{Q} + \KL{P}{Q}, \IEEEeqnarraynumspace 
\end{IEEEeqnarray}
where~\eqref{EqDecember22at19h48in2024C} follows from~\eqref{December15at16h31in2024}; and  ~\eqref{EqDecember22at19h48in2024D} follows from~\eqref{EqDecember19at14h35in2024Nice}.  
\end{IEEEproof}
It is interesting to highlight  that~$\mathsf{G}_h\left( x, P, P^{(h, Q, \lambda)}_{Y | X = x } \right)$ in~\eqref{EqDecember20at17h39in2024BusFromSophiaToNice} characterizes the variation of the expectation of the function~$h(x,\cdot): \reals^{m} \to \reals$, when $\lambda > 0$ (resp. $\lambda < 0$) and  the probability measure changes from the solution to the optimization problem~\eqref{EqDecember18at11h09in2024Sophia} (resp.~\eqref{EqDecember18at11h09in2024SophiaA}) to an alternative measure~$P$.

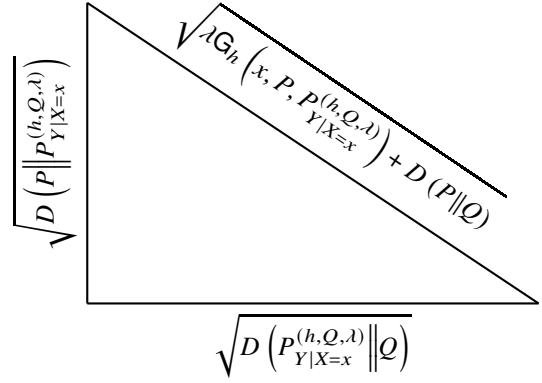
\begin{figure}[t]
\begin{center}
\begin{tikzpicture}
\draw[thick] (0,0) -- node[below] {$\sqrt{ \KL{P^{(h, Q, \lambda)}_{Y | X = x }}{Q} }$} ++(6,0);
\draw[thick] (0,0) -- node[left,rotate=90, above] {$\sqrt{\KL{P}{P^{(h, Q, \lambda)}_{Y | X = x }}}$} ++(0,4);
\draw[thick] (0,4) -- (6,0);
\node[left, rotate=-34, above] at (3,2) {$\sqrt{\lambda \mathsf{G}_h\left( x, P, P^{(h, Q, \lambda)}_{Y | X = x } \right) + \KL{P}{Q}}$};
\end{tikzpicture}
\end{center}
\caption{Geometric interpretation of  Lemma~\ref{LemmaDecember20at17h39in2024BusFromSophiaToNice}, with $Q$ a probability measure.}
\label{FigSeptember4at18h58in2024}
\end{figure}

Finally, when the reference measure $Q$ is a probability measure, the converse of  the Pythagorean theorem~\cite[Book~I, Proposition~$48$]{heath1956thirteen} together with  Lemma~\ref{LemmaDecember20at17h39in2024BusFromSophiaToNice}, lead to the geometric construction shown in Figure~\ref{FigSeptember4at18h58in2024}. 
Such a geometric interpretation is similar to those presented in~\cite[Figure 6]{perlaza2024generalization} and~\cite[Figure 7]{perlaza2024generalization} in the context of the generalization error of machine learning algorithms. The former considers $\lambda>0$, while the latter considers $\lambda < 0$. 
Nonetheless, the interpretation in Figure~\ref{FigSeptember4at18h58in2024} is general and independent of such an application.

\section{Characterization of~$\mathsf{G}_{h}\left( x, P_1, P_2 \right)$ in~\eqref{EqDecember11at20h03in2024AgadirCity}}
\label{sec:ii}
The main result of this section is the following theorem.
\begin{theorem}\label{TheoremDecember15at12h11in2024Nice}
For all probability measures~$P_{1}$ and~$P_{2}$, both absolutely continuous with respect to a given $\sigma$-finite measure~$Q$ on $\reals^{m}$, the variation~$\mathsf{G}_{h}\left( x, P_1, P_2 \right)$ in~\eqref{EqDecember11at20h03in2024AgadirCity} satisfies, 
\begin{IEEEeqnarray}{rCl}
\nonumber
\mathsf{G}_{h}\left( x, P_1, P_2 \right) & = & \frac{1}{\lambda}\bigg( \KL{P_1}{P^{\left(h, Q, \lambda\right)}_{Y|X = x}}  - \KL{P_2}{P^{\left(h, Q, \lambda\right)}_{Y|X = x}}  
\\ 
\label{EqDecember15at12h03in2024Nice}
& & + \KL{P_2}{Q} - \KL{P_1}{Q} \bigg), 
\end{IEEEeqnarray}
where the probability measure~$P^{\left(h, Q, \lambda\right)}_{Y|X = x}$, with~$\lambda \neq 0$, is an~$(h, Q, \lambda)$-Gibbs probability measure.
\end{theorem}
\begin{IEEEproof}
The proof follows from Lemma~\ref{LemmaDecember20at17h39in2024BusFromSophiaToNice} and by observing that~$\mathsf{G}_{h}\left( x, P_1, P_2 \right) =  \mathsf{G}_h\left( x, P_1, P^{(h, Q, \lambda)}_{Y | X = x } \right) - \mathsf{G}_h\left( x, P_2, P^{(h, Q, \lambda)}_{Y | X = x } \right)$.
\end{IEEEproof}

Theorem~\ref{TheoremDecember15at12h11in2024Nice} might be particularly simplified in the case in which the reference measure~$Q$ is a probability measure. 
Consider for instance the case in which~$P_1\ll P_{2}$ (or~$P_{2}\ll P_{1}$). In such a case, the reference measure might be chosen as~$P_2$ (or~$P_1$), as shown hereunder.
\begin{corollary}\label{CorollaryDecember17at16h17n2024BusFromSophiaToNice}
Consider the variation~$\mathsf{G}_{h}\left( x, P_1, P_2 \right)$ in~\eqref{EqDecember11at20h03in2024AgadirCity}. If the probability measure~$P_1$ is absolutely continuous with respect to~$P_{2}$, then, 
\begin{IEEEeqnarray}{rCl}
\nonumber
\mathsf{G}_{h}\left( x, P_1, P_2 \right) & = &  \frac{1}{\lambda} \bigg( \KL{P_1}{P^{\left(h, P_2, \lambda\right)}_{Y|X = x}}  - \KL{P_2}{P^{\left(h, P_2, \lambda\right)}_{Y|X = x}} 
\\ 
\label{EqDecember17at16h25in2024BusFromSophiaToNice}
& &  - \KL{P_1}{P_2} \bigg). 
\end{IEEEeqnarray}
Alternatively, if the probability measure~$P_2$ is absolutely continuous with respect to~$P_{1}$, then, 
\begin{IEEEeqnarray}{rCl}
\nonumber
\mathsf{G}_{h}\left( x, P_1, P_2 \right) & = &  \frac{1}{\lambda}\bigg( \KL{P_1}{P^{\left(h, P_1, \lambda\right)}_{Y|X = x}}  - \KL{P_2}{P^{\left(h, P_1, \lambda\right)}_{Y|X = x}}  
\\ 
\label{EqDecember17at16h25in2024BusFromSophiaToNiceB}
& &  + \KL{P_2}{P_1} \bigg),
\end{IEEEeqnarray}
where the probability measures~$P^{\left(h, P_1, \lambda\right)}_{Y|X = x}$ and~$P^{\left(h, P_2, \lambda\right)}_{Y|X = x}$  are respectively~$(h, P_1, \lambda)$- and~$(h, P_2, \lambda)$-Gibbs probability measures, with~$\lambda \neq 0$. 
\end{corollary}
In the case in which neither~$P_1$ is absolutely continuous with respect to~$P_{2}$; nor~$P_2$ is absolutely continuous with respect to~$P_{1}$, the reference measure~$Q$ in Theorem~\ref{TheoremDecember15at12h11in2024Nice} can always be chosen as a convex combination of~$P_1$ and~$P_2$. That is, for all Borel sets~$\set{A} \in \BorSigma{\reals^{m}}$, 
$Q \left( \set{A} \right) = \alpha P_{1} \left( \set{A} \right) + (1 - \lambda) P_{2} \left( \set{A} \right)$, with~$\alpha \in (0,1)$.

Theorem~\ref{TheoremDecember15at12h11in2024Nice} can be especialized to the
specific cases in which $Q$ is the Lebesgue or the counting measure.

\paragraph{If $Q$ is the Lebesgue measure}
the probability measures~$P_1$ and~$P_2$ in~\eqref{EqDecember15at12h03in2024Nice} admit probability density functions~$f_1$ and~$f_2$, respectively. Moreover, the terms~$-\KL{P_1}{Q}$ and~$-\KL{P_2}{Q}$ are Shannon's differential entropies \cite{Shannon-1948a} induced by~$P_1$ and~$P_2$, denoted by~$\mathsf{h}(P_1)$ and~$\mathsf{h}(P_2)$, respectively. That is, for all~$i \in \lbrace 1,2 \rbrace$,
%
$\mathsf{h}(P_i)  \triangleq   - \int f_{i}(x) \log f_{i}(x) \mathrm{d}x$.
%
The probability measure~$P^{\left(h, Q, \lambda\right)}_{Y|X = x}$, with~$\lambda \neq 0$,~$x \in \reals^{n}$, and~$Q$ the Lebesgue measure, possesses a  probability density function, denoted by~$f^{\left(h, Q, \lambda\right)}_{Y|X = x}: \reals^m \to (0, +\infty)$, which  satisfies 
\begin{IEEEeqnarray}{rcl}
\label{EqDecember23at14h36in2024Nice}
f^{\left(h, Q, \lambda\right)}_{Y|X = x}(y) & = & \frac{\exp\left( -\lambda h(x,y) \right)}{\int \exp\left( -\lambda h(x,y) \right) \mathrm{d}y}.
\end{IEEEeqnarray}

\paragraph{If $Q$ is the counting measure}
the probability measures~$P_1$ and~$P_2$ in~\eqref{EqDecember15at12h03in2024Nice} admit probability mass functions~$p_1:\set{Y} \to [0,1]$  and~$p_2:\set{Y} \to [0,1]$, with~$\set{Y}$ a countable subset of~$\reals^{m}$. Moreover,~$-\KL{P_1}{Q}$ and~$-\KL{P_2}{Q}$ are respectively  Shannon's discrete entropies~\cite{Shannon-1948a} induced by~$P_1$ and~$P_2$, denoted by~$\mathsf{H}(P_1)$ and~$\mathsf{H}(P_2)$, respectively. 
That is, for all~$i \in \lbrace 1,2 \rbrace$,
\begin{IEEEeqnarray}{rCl}
\label{EqDecember24at11h24in2024Nice}
\mathsf{H}(P_i) & \triangleq & - \sum_{y \in \set{Y}} p_{i}(y) \log p_{i}(y). 
\end{IEEEeqnarray}
The probability measure~$P^{\left(h, Q, \lambda\right)}_{Y|X = x}$, with~$\lambda \neq 0$ and~$Q$ the counting measure, possesses a conditional probability mass function, denoted by~$p^{\left(h, Q, \lambda\right)}_{Y|X = x}: \set{Y} \to (0, +\infty)$,   which  satisfies 
\begin{IEEEeqnarray}{rcl}
\label{EqDecember23at14h35in2024Nice}
p^{\left(h, Q, \lambda\right)}_{Y|X = x}(y) & = & \frac{\exp\left( -\lambda h(x,y) \right)}{\sum_{y \in \set{Y}} \exp\left( -\lambda h(x,y) \right)}.
\end{IEEEeqnarray}
\section{Characterizations of~$\bar{\mathsf{G}}_{h}\left( P_{Y|X}^{(1)}, P_{Y|X}^{(2)}, P_{X} \right)$ in~\eqref{EqDecember11at20h13in2024AgadirCity}}
\label{sec:iii}

The main result of this section is a characterization of~$\bar{\mathsf{G}}_{h}\left( P_{Y|X}^{(1)}, P_{Y|X}^{(2)}, P_{X} \right)$ in~\eqref{EqDecember11at20h13in2024AgadirCity}. 
\begin{theorem}\label{TheoremDecember23at13h22in2024Nice}
Consider the variation~$\bar{\mathsf{G}}_h\left( P_{Y|X}^{(1)}, P_{Y|X}^{(2)}, P_{X} \right)$ in~\eqref{EqDecember11at20h13in2024AgadirCity} and assume that for all~$x \in \reals^{n}$, the probability measures~$P_{Y|X = x}^{(1)}$ and~$P_{Y|X = x}^{(2)}$ are both absolutely continuous with respect to a~$\sigma$-measure~$Q$. Then, 
\begin{IEEEeqnarray}{rCl}
\nonumber
& & \bar{\mathsf{G}}_h\left( P_{Y|X}^{(1)}, P_{Y|X}^{(2)}, P_{X} \right)  \\
\nonumber
& = & \frac{1}{\lambda} \int \bigg( \KL{P_{Y|X = x}^{(1)}}{P^{\left(h, Q, \lambda\right)}_{Y|X = x}}  - \KL{P_{Y|X = x}^{(2)}}{P^{\left(h, Q, \lambda\right)}_{Y|X = x}} 
\\ 
\label{EqDecember26at12h48in2024Nice}
& & + \KL{P_{Y|X = x}^{(2)}}{Q} - \KL{P_{Y|X=x}^{(1)}}{Q} \bigg) \mathrm{d}P_{X}(x), 
\end{IEEEeqnarray}
where the probability measure~$P^{\left(h, Q, \lambda\right)}_{Y|X}$, with~$\lambda \neq 0$, is an~$(h, Q, \lambda)$-Gibbs conditional probability measure.
\end{theorem}
\begin{IEEEproof}
The proof follows from~\eqref{EqDecember11at20h13in2024AgadirCity} and Theorem~\ref{TheoremDecember15at12h11in2024Nice}.
\end{IEEEproof}

Note that, from~\eqref{EqDecember11at20h13in2024AgadirCity}, it follows that the general expression for the expected variation~$\bar{\mathsf{G}}_h\left( P_{Y|X}^{(1)}, P_{Y|X}^{(2)}, P_{X} \right)$ might be simplified according to Corollary~\ref{CorollaryDecember17at16h17n2024BusFromSophiaToNice}. 
For instance,  if for all~$x \in \reals^{m}$, the probability measure~$P_{Y|X=x}^{(1)}$ is absolutely continuous with respect to~$P_{Y|X = x}^{(2)}$, the measure~$P_{Y|X = x}^{(2)}$ can be chosen to be the reference measure in the calculation of~$\mathsf{G}_h\left( x, P_{Y|X=x}^{(1)}, P_{Y|X = x}^{(2)} \right)$ in~\eqref{EqDecember11at20h13in2024AgadirCity}. This observation leads to the following corollary of Theorem~\ref{TheoremDecember23at13h22in2024Nice}.
\begin{corollary}\label{CorDecember24at9h07in2024}
Consider the variation~$\bar{\mathsf{G}}_h\left( P_{Y|X}^{(1)}, P_{Y|X}^{(2)}, P_{X} \right)$ in~\eqref{EqDecember11at20h13in2024AgadirCity} and assume that for all~$x \in \reals^{n}$,~$P_{Y|X = x}^{(1)}\ll P_{Y|X = x}^{(2)}$.~Then, 
\begin{IEEEeqnarray}{lCl}
\nonumber
& & \bar{\mathsf{G}}_h\left( P_{Y|X}^{(1)}, P_{Y|X}^{(2)}, P_{X} \right)  \\
\nonumber
& = & \frac{1}{\lambda} \int \Bigg( \KL{P_{Y|X = x}^{(1)}}{P^{\left(h, P_{Y|X = x}^{(2)}, \lambda\right)}_{Y|X = x}} - \KL{P_{Y|X = x}^{(2)}}{P^{\left(h, P_{Y|X = x}^{(2)}, \lambda\right)}_{Y|X = x}} \squeezeequ\\
& &   - \KL{P_{Y|X=x}^{(1)}}{P_{Y|X = x}^{(2)}} \Bigg) \mathrm{d}P_{X}(x).  \IEEEeqnarraynumspace
\end{IEEEeqnarray}
Alternatively,  if for all~$x \in \reals^{n}$, the probability measure~$P_{Y|X = x}^{(2)}$ is absolutely continuous with respect to~$P_{Y|X = x}^{(1)}$, then, 
\begin{IEEEeqnarray}{lCl}
\nonumber
\bar{\mathsf{G}}_h\left( P_{Y|X}^{(1)}, P_{Y|X}^{(2)}, P_{X} \right)  =  \frac{1}{\lambda} \int \Bigg( \KL{P_{Y|X = x}^{(1)}}{P^{\left(h, P_{Y|X=x}^{(1)}, \lambda\right)}_{Y|X = x}} \squeezeequ \\
 - \KL{P_{Y|X = x}^{(2)}}{P^{\left(h, P_{Y|X=x}^{(1)}, \lambda\right)}_{Y|X = x}}  + \KL{P_{Y|X = x}^{(2)}}{P_{Y|X = x}^{(1)}} \Bigg) \mathrm{d}P_{X}(x), \middlesqueezeequ  \IEEEeqnarraynumspace 
\end{IEEEeqnarray}
where the measures~$P^{\left(h, P_{Y|X = x}^{(1)}, \lambda\right)}_{Y|X = x}$ and~$P^{\left(h, P_{Y|X = x}^{(2)}, \lambda\right)}_{Y|X = x}$  are~$(h, P_{Y|X = x}^{(1)}, \lambda)$- and~$(h, P_{Y|X = x}^{(2)}, \lambda)$-Gibbs probability measures, respectively. 
\end{corollary}
The Gibbs probability measures~$P^{\left(h, P_{Y|X=x}^{(1)}, \lambda\right)}_{Y|X = x}$ and~$P^{\left(h, P_{Y|X = x}^{(2)}, \lambda\right)}_{Y|X = x}$ in Corollary~\ref{CorDecember24at9h07in2024} are particularly interesting as their reference measures depend on~$x$.  Gibbs measures of this form appear, for instance, in \cite[Corollary~$10$]{PerlazaTIT2024}.

Two special cases are particularly noteworthy.

\paragraph{When the reference measure~$Q$ is the Lebesgue measure} observe that the terms~$-\int \KL{P_{Y|X = x}^{(1)}}{Q} \mathrm{d}P_{X}(x)$ and~$-\int \KL{P_{Y|X=x}^{(2)}}{Q} \mathrm{d}P_{X}(x)$ in~\eqref{EqDecember26at12h48in2024Nice} both become Shannon's differential conditional entropy, denoted by $\mathsf{h} \left( P_{Y|X}^{(1)} | P_{X}\right)$ and $\mathsf{h} \left( P_{Y|X}^{(2)} | P_{X}\right)$, respectively.  That is, for all~$i \in \lbrace 1, 2 \rbrace$,
$\mathsf{h} \left( P_{Y|X}^{(i)} | P_{X}\right) \triangleq  \int \mathsf{h}\left( P_{Y|X = x}^{(i)}\right) \mathrm{d}P_{X}(x),$
where~$\mathsf{h}$ is the entropy functional in~\eqref{EqDecember24at11h23in2024Nice}.

\paragraph{When the reference measure~$Q$ is the counting measure} the terms~$-\int \KL{P_{Y|X = x}^{(1)}}{Q} \mathrm{d}P_{X}(x)$ and~$-\int \KL{P_{Y|X=x}^{(2)}}{Q} \mathrm{d}P_{X}(x)$ in~\eqref{EqDecember26at12h48in2024Nice} both become Shannon's discrete conditional entropies, denoted by $\mathsf{H} \left( P_{Y|X}^{(1)} | P_{X}\right)$ and $\mathsf{H} \left( P_{Y|X}^{(2)} | P_{X}\right)$, respectively.  That is, for all~$i \in \lbrace 1, 2 \rbrace$,
\begin{IEEEeqnarray}{rcl}
\mathsf{H} \left( P_{Y|X}^{(i)} | P_{X}\right) & \triangleq & \int \mathsf{H}\left( P_{Y|X = x}^{(i)}\right) \mathrm{d}P_{X}(x),
\end{IEEEeqnarray}
where~$\mathsf{H}$ is the entropy functional in~\eqref{EqDecember24at11h24in2024Nice}.

\section{Characterizations of~$\bar{\mathsf{G}}_{h}\left( P_{Y}, P_{Y|X}, P_{X} \right)$ in \eqref{EqDecember24at12h06in2024Nice}}

The main result of this section is a characterization of~$\bar{\mathsf{G}}_{h}\left( P_{Y}, P_{Y|X}, P_{X} \right)$ in~\eqref{EqDecember24at12h06in2024Nice}, which describes the variation of the expectation of the function $h$ when the probability measure changes from the joint probability measure $P_{Y|X}P_{X}$ to the product of its marginals $P_{Y} \cdot P_{X}$. 
This result is presented hereunder and involves the mutual information $I\left( P_{Y|X} ; P_{X}\right)$ and lautum information $L\left( P_{Y|X} ; P_{X}\right)$, defined as follows:
\begin{IEEEeqnarray}{rcl}
I\left( P_{Y|X} ; P_{X}\right) & \triangleq & \int \KL{P_{Y|X = x}}{P_{Y}} \mathrm{d}P_{X}(x); \mbox{ and }\\
L\left( P_{Y|X} ; P_{X}\right) & \triangleq & \int \KL{P_{Y}}{P_{Y|X = x}} \mathrm{d}P_{X}(x).
\end{IEEEeqnarray}

\begin{theorem}\label{TheoremDecember25at20h12in2024Nice}
Consider the expected variation $\bar{\mathsf{G}}_h\left( P_{Y}, P_{Y|X}, P_{X} \right)$ in~\eqref{EqDecember24at12h06in2024Nice} and assume that, for all $x \in \reals^{n}$:
\begin{enumerate}[(a)]
\item The probability measures $P_Y$ and $P_{Y|X = x}$ are both absolutely continuous with respect to a given $\sigma$-finite measure $Q$; and 
\item The probability measures $P_Y$ and $P_{Y|X = x}$ are mutually absolutely continuous.
\end{enumerate}
Then, it follows that
\begin{IEEEeqnarray}{rcl}
\nonumber
& & \bar{\mathsf{G}}_h\left( P_{Y}, P_{Y|X}, P_{X} \right) = \frac{1}{\lambda}\Bigg( I\left( P_{Y|X} ; P_{X}\right) +  L\left( P_{Y|X} ; P_{X}\right) \\
\nonumber
& & + \int \int \log\left( \frac{\mathrm{d}P_{Y|X = x}}{\mathrm{d}P^{\left(h, Q, \lambda\right)}_{Y|X = x}} (y)\right) \mathrm{d} P_{Y}(y)\mathrm{d} P_{X}(x) \\
\label{EqDecember24at17h55Nice}
& & - \int \int \log\left( \frac{\mathrm{d}P_{Y|X = x}}{\mathrm{d}P^{\left(h, Q, \lambda\right)}_{Y|X = x}} (y)\right) \mathrm{d} P_{Y|X = x}(y)\mathrm{d} P_{X}(x) \Bigg), \IEEEeqnarraynumspace
\end{IEEEeqnarray}
where the probability measure~$P^{\left(h, Q, \lambda\right)}_{Y|X}$, with~$\lambda \neq 0$, is an~$(h, Q, \lambda)$-Gibbs conditional probability measure.
\end{theorem}
\begin{IEEEproof}
The proof follows from Theorem~\ref{TheoremDecember23at13h22in2024Nice}, which holds under assumption $(a)$ and leads to
\begin{IEEEeqnarray}{rcl}
\nonumber
& & \bar{\mathsf{G}}_h\left( P_{Y}, P_{Y|X}, P_{X} \right)  \\
\nonumber
& = & \frac{1}{\lambda} \int \bigg( \KL{P_{Y}}{P^{\left(h, Q, \lambda\right)}_{Y|X = x}}  - \KL{P_{Y|X = x}}{P^{\left(h, Q, \lambda\right)}_{Y|X = x}} 
\\ 
\label{EqDecember24at13h57in2024Nice}
& & + \KL{P_{Y|X=x}}{Q} - \KL{P_{Y}}{Q} \bigg) \mathrm{d}P_{X}(x).
\end{IEEEeqnarray}
The proof continues by noticing that  
\begin{IEEEeqnarray}{rcl}
\label{EqDecember24at13h57in2024NiceI}
\int \KL{P_{Y|X=x}}{Q} \mathrm{d}P_{X}(x)  & = & I\left( P_{Y|X} ; P_{X}\right)  + \KL{P_Y}{Q},\IEEEeqnarraynumspace
 \end{IEEEeqnarray}
and %
\begin{IEEEeqnarray}{rcl}
\nonumber
& &\int \KL{P_{Y}}{P^{\left(h, Q, \lambda\right)}_{Y|X = x}}  \mathrm{d}P_{X}(x)  =  L\left( P_{Y|X} ; P_{X}\right) \\
\label{EqDecember24at17h46in2024Nice}
& & + \int \int \log\left( \frac{\mathrm{d}P_{Y|X = x}}{\mathrm{d}P^{\left(h, Q, \lambda\right)}_{Y|X = x}} (y)\right) \mathrm{d} P_{Y}(y)\mathrm{d} P_{X}(x).
\end{IEEEeqnarray}
Finally, using \eqref{EqDecember24at13h57in2024NiceI} and \eqref{EqDecember24at17h46in2024Nice} in \eqref{EqDecember24at13h57in2024Nice} yields \eqref{EqDecember24at17h55Nice}, which completes the proof.
\end{IEEEproof}

An interesting observation from Theorem~\ref{TheoremDecember25at20h12in2024Nice} is that the last two terms in the right-hand side of \eqref{EqDecember24at17h55Nice} are both zero in the case in which $P_{Y|X}$ is an~$(h, Q, \lambda)$-Gibbs conditional probability measure. This is observation is highlighted by the following corollary.
\begin{corollary}
Consider an~$(h, Q, \lambda)$-Gibbs conditional probability measure, denoted by~$P^{(h, Q, \lambda)}_{Y | X} \in \triangle\left( \reals^{m} | \reals^{n} \right)$, with~$\lambda \neq 0$; and a probability measure $P_{X} \in \triangle\left( \reals^{n} \right)$. Let the measure $P^{(h, Q, \lambda)}_{Y} \in \triangle\left( \reals^{m} \right)$ be such that for all sets $\set{A} \in \BorSigma{\reals^{m}}$, 
\begin{IEEEeqnarray}{rcl}
 P^{(h, Q, \lambda)}_{Y} \left( \set{A} \right) & = & \int P^{(h, Q, \lambda)}_{Y | X = x}  \left( \set{A} \right) \mathrm{d} P_{X}(x).
\end{IEEEeqnarray}
Then, 
\begin{IEEEeqnarray}{rcl}
\nonumber
& & \bar{\mathsf{G}}_h\left( P^{(h, Q, \lambda)}_{Y}, P^{(h, Q, \lambda)}_{Y | X}, P_{X} \right) \\
& = &\frac{1}{\lambda} \left( I\left(P^{(h, Q, \lambda)}_{Y | X} ; P_{X} \right) +  L\left( P^{(h, Q, \lambda)}_{Y | X} ; P_{X} \right) \right).
\end{IEEEeqnarray}
\end{corollary}
\vspace{-1ex}
\IEEEtriggeratref{10}
\bibliographystyle{IEEEtran}
\bibliography{ReferencesISIT2025}
\end{document}